\documentclass[twocolumn,showpacs,preprintnumbers,amsmath,amssymb]{revtex4}
\usepackage{graphicx}% Include figure files
\usepackage{epstopdf}
\usepackage{amssymb}
\usepackage{dcolumn}% Align table columns on decimal point
\usepackage{bm}% bold math

\begin{document}

  \title{Is the coexistence of Catalan and Spanish possible in Catalonia?}

  \author{Lu\'is F. Seoane$^{1,2,3,@,*}$, Xaqu\'in Loredo$^{4}$, Henrique Monteagudo$^{5,6}$, and Jorge Mira$^{7,@,*}$}

  \affiliation{ 
      \\ $^1$ Department of Physics, Massachusetts Institute of Technology, Cambridge, MA 02139. 
      \\ $^2$ ICREA-Complex Systems Lab, Universitat Pompeu Fabra -- PRBB, Dr. Aiguader 88, 08003 Barcelona, Spain. 
      \\ $^3$ Institut de Biologia Evolutiva, UPF-CSIC, Passg Barceloneta, 08003 Barcelona. 
      \\ $^4$ Seminario de Socioling\"u\'istica da Real Academia Galega, Santiago de Compostela, Spain. 
      \\ $^5$ Real Academia Galega and Instituto da Lingua Galega. 
      \\ $^6$ Universidade de Santiago de Compostela, 15782 Santiago de Compostela, Spain. 
      \\ $^7$ Departamento de F\'isica Aplicada, Universidade de Santiago de Compostela, 15782 Santiago de Compostela, Spain. 
      \\ $^@$ seoane@mit.edu, jorge.mira@usc.es. }

  \begin{abstract}

    We study the stability of two coexisting languages (Catalan and Spanish) in Catalonia (North-Eastern Spain), a key European region in political and economic terms. Our analysis relies on recent, abundant empirical data that is studied within an analytic model of population dynamics. This model contemplates the possibilities of long-term language coexistence or extinction. We establish that the most likely scenario is a sustained coexistence. The data needs to be interpreted under different circumstances, some of them leading to the asymptotic extinction of one of the languages involved. We delimit the cases in which this can happen. Asymptotic behavior is often unrealistic as a predictor for complex social systems, hence we make an attempt at forecasting trends of speakers towards $2030$. These also suggest sustained coexistence between both tongues, but some counterintuitive dynamics are unveiled for extreme cases in which Catalan would be likely to lose an important fraction of speakers. As an intermediate step, model parameters are obtained that convey relevant information about the prestige and interlinguistic similarity of the tongues as perceived by the population. This is the first time that these parameters are quantified rigorously for this couple of languages. Remarkably, Spanish is found to have a larger prestige specially in areas which historically had larger communities of Catalan monolingual speakers. Limited, spatially-segregated data allows us to examine more fine grained dynamics, thus better addressing the likely coexistence or extinction. Variation of the model parameters across regions are informative about how the two languages are perceived in more urban or rural environments.

  \end{abstract}


\maketitle

  \section{Introduction}
  	\label{sec:1}

    Catalan is a Romance language evolved in the north-east of the Iberian peninsula after the fall of the Roman Empire \cite{Lleal1990, Vila2008}. This language was linked to the political institutions emerging in that region within the Crown of Arag\'on. This polity included some of the south departments of modern France and the prominent county of Barcelona, often center of the government of all these territories. The Crown of Arag\'on expanded southwards along the Mediterranean coastline (nowadays Valencia) and beyond the Iberian peninsula bringing the Catalan language to the Balearic and other Mediterranean islands. Variants of this tongue are still spoken in Sardinia and southern France. 

    Castillian (nowadays Spanish) is a Romance language evolved in the north central part of the Iberian peninsula at the same time as Catalan \cite{Lleal1990}. Castillian was linked to the political institutions of the region, notably the Crown of Castille, which extended southwards, like Arag\'on, as the struggle between the Christian and Islamic forces unfolded in the Iberian peninsula during the Middle Ages. Similar historic stages can be seen in the spread of Portuguese along the Atlantic coast, while Castillian advanced across the inland territories. When the colonial era began, both Castillian and Portuguese diffused prominently through the American continent \cite{Penny2002, Pharies2008}. 

    Back in the Iberian Peninsula, the Portuguese and Castillian kingdoms remained different political entities in the long term. The Crowns of Arag\'on and Castille became unified forming the embryo of modern Spain (which involved also other regional kingdoms). Explicit political measurements were tried to mitigate the social split between the populations originating from the old, separate kingdoms. But, despite these efforts, the Catalan-Castillian linguistic division lasted through the creation of the Spanish nation-state. The existence of a vernacular language (Catalan) different from the official language of the Spanish kingdom (Castillian or Spanish), together with other cultural singularities, provided support for the emergence of regionalist and nationalist movements during the XIX and XX centuries in Catalonia and other Catalan-speaking regions \cite{Llobera2003}. After the Spanish Civil war (1936-1939), Catalan was repressed and banned from official communications \cite{Guibernau2004}. The restoration of democracy in Spain (symbolized by its modern constitution from 1978) awarded Catalan a co-official status in Catalonia. Different policies have ever since been articulated to protect Catalan from decline -- these range from subsidized presence in the mass media to different strategies within the educative systems \cite{Pradilla2001}. 

    The past and future evolution of the Spanish and Catalan languages is tightly linked to the complex political scenario of modern Spain. Secessionist Catalan movements have grown steadily during the past decades \cite{growingIndep}, currently reaching a climax with an explicit agenda towards full independence formulated by the Catalan government. While language use is not a definitive factor, it certainly correlates with the different political actors \cite{Clua2014,Woolard2008}. Understanding and forecasting the evolution of the Catalan-Spanish tongues in Catalonia from a population dynamics perspective is, hence, a relevant political goal. 

    These dynamics become even more important from a sociolinguistic point of view. We worry about the survival and peaceful coexistence of linguistic communities. This is the mindset that we wish to adopt in this paper. Our approach is of relevance beyond the specific example that we deal with here. The total number of languages across the globe is reported to decline and the number of endangered tongues augments \cite{Sutherland2003}. While this is the overall context, here we focus on the particular Catalan-Spanish coexisting dynamics. Abundant, curated data is available for this system. Also, as argued above, the ongoing political scenario makes it a very appealing study case. \\

      \begin{figure*}
        \centerline{\includegraphics[width=\linewidth]{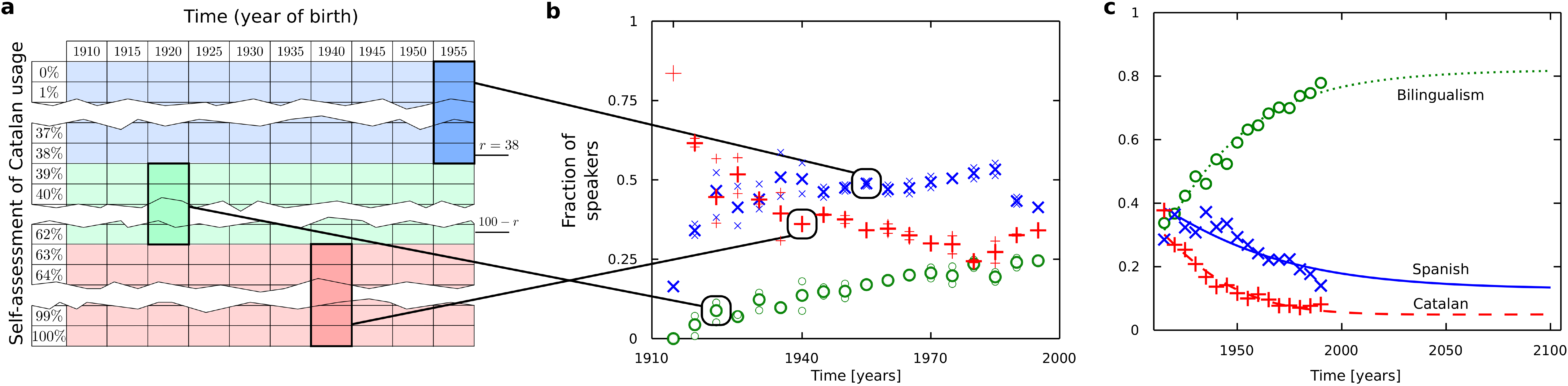}}
        \caption{{\bf From data to theory. } {\bf a} A sample of Catalan citizens classified their daily usage of Catalan within $0-100\%$ (correspondingly, of Spanish within $100-0\%$). By setting a bilingualism threshold $r$, we accept as bilinguals those individuals using both languages a percentage greater than $r$ (in this figure we chose $r=38$ arbitrarily). They occupy the central rows in the table in panel {\bf a}. Speakers declaring to use one of the languages a percentage less than $r$ were classified as monolinguals (top rows of the table for Spanish monolinguals, with Catalan usage below $38\%$; and bottom rows for Catalan monolinguals). {\bf b} Speakers have also been aggregated in five-year intervals according to their year of birth. With the fractions $x$ (blue crosses, Spanish monolinguals), $b$ (green circles, bilinguals) and $y$ (red plus symbols, Catalan monolinguals) of speakers in each linguistic group and their average years of birth we reconstruct time series that we will fit to our model equations. Two different surveys are available (each of these series is represented by smaller symbols in panel {\bf b}). We can average these data (larger symbols) to create more robust estimators of the fractions of speakers. For each $r$ a different data series is created. Instead of considering a rigid definition of bilingualism (whose existence we do not debate), we conducted our analysis with all possible integer prescriptions ($r \in [1, 50]$). {\bf c} Example fit to the data for $r=10$. It predicts a majority of bilingual speakers in the long term with small, similarly sized monolingual groups. This is consistent with the broad definition of bilingualism that $r=10$ implies. }\label{fig:01}
      \end{figure*}

    The mathematical characterization of population dynamics is well rooted within the field of ecology \cite{Kot2001, Turching2003}. Recent works have extended this kind of analysis to different social aspects, including the evolution of speakers of different coexisting languages \cite{BaggsFreedman1990, BaggsFreedman1993, AbramsStrogatz2003, MiraParedes2005, Kandler2008, Castellano2009, PatriarcaHeinsalu2009, CastelloSan2013, ZhangGong2013}. The approach is based on sets of differential equations able to reconstruct historical series of data and, hopefully, make informative predictions. 
    
    The reconstruction part of this problem has been successful in various scenarios. Relevant examples are the modeling of up to $42$ cases of language coexistence by Abrams and Strogatz \cite{AbramsStrogatz2003}. One of the studied cases was that of Sottish Gaelic. This was further investigated in its full complexity (which implies language dynamics across different territories) by Kandler et al. \cite{KandlerSteele2010}. Besides accounting for historical data, these authors make a first effort in prediction in an uncertain political environment. This is a scenario similar to ours. Forecasting in the face of unsettled (political) struggles is an ambitious goal that also calls for a warning: there is a limit to the contingencies that our mathematical models can account for, and hence no prediction can be taken as definitive. This is also an opportunity for the valuable interplay between theory, its predictions, and its prospective failures; to improve our models. 

    We based our work on a vast amount of official data collected by the Institut d'Estad\'istica de Catalunya (Catalan Institute of Statistics) during the last decades \cite{EULP2003a, EULP2003b, EULP2008a, EULP2008b, EULPSummary, EULP2013}. These are very rich data sets with more details available than our models can account for. Hence, important simplifications and preprocessing of the data (described below) were necessary. Ultimately, we modeled the data according to the proposal by Mira et al. \cite{MiraParedes2005}. This relies on a system of differential equations that track over time two monolingual populations along a bilingual one. The stability of this model has been characterized during the last years \cite{MiraNieto2011, OteroMira2013, ColucciOtero2014, SeoaneMira2017}, which makes it a powerful tool for our analysis. However, we wish to invite future contributions to examine the same or similar data using alternative equations as well.

  \section{Methods}
    \label{sec:2}

    \subsection{Data sources and pre-processing}

      The original data were gathered by the Institut d'Estad\'istica de Catalunya (Idescat, {\em Catalan Institute of Statistics}) in the EULP (Enquesta d'Usos Ling\"u\'istics de la Poblaci\'o, {\em Survey of Language Use by the Population}) surveys \cite{EULP2003a, EULP2003b, EULP2008a, EULP2008b, EULPSummary, EULP2013}, which have been conducted every five years since $2003$. The relevant item for us is the self-assessment of language use, in which the respondents reported as a percentage their daily use of each language as detailed below. This item is absent in the $2003$ studies \cite{EULP2003a, EULP2003b}, so we only used the $2008$ and $2013$ editions \cite{EULP2008a, EULP2008b, EULP2013}. 

      Each speaker self-assessed, as a percentage, her daily use of each of the tongues that she speaks. These included Catalan and Spanish together with Galician, Arab, Urdu, and many others arising from different migratory currents. We want to focus on the Catalan-Spanish dynamics alone. We removed all speakers that used any other language more than $30\%$ of the time. For the speakers retained, we discarded the other languages reported and normalized the data such that Catalan plus Spanish add up to $100\%$. The respondents were stratified in $5$-year intervals to generate a table (Fig. \ref{fig:01}{\bf a}) in which each square is associated to the average date in which respondents were born (spanning from $1910$ to $1990$ in EULP-2008 and from $1915$ to $1995$ in EULP-2013) and the percentage of Catalan use that they reported. Hence, each entry of the table contains the estimated number of people in Catalonia of the same age that would report a same percentage of Catalan use. We treat this age-stratified data as a proxy about the proportion of Catalan speakers at the time that each group was born. This is inspired by other {\em apparent-time} studies \cite{Labov1963, Eckert1997,Mague2006, Chambers2013}. This approximation is not free of criticism and has been more often used to study variants of a same language, but it is a suitable way to generate a data series from the available data. 

      In Catalonia more than half of the population concentrates in Barcelona and its metropolitan area, which may arguably have different dynamics from the rest of the region. Accordingly, after studying the population dynamics inferred from global time series, we repeated the analysis for these two segregated regions (Barcelona plus metropolitan area vs rest of the territory). 

      In every case, a table similar to that in Fig. \ref{fig:01}{\bf a} constitutes our raw data. Several models discussed in the literature \cite{BaggsFreedman1990, BaggsFreedman1993, MiraParedes2005, ZhangGong2013, MinettWang2008, HeinsaluLeonard2014} coarse grain language use into three broad categories: two monolinguals and a bilingual one. The model chosen for our analysis \cite{MiraParedes2005} (see below) does so. We imposed these divisions on our raw data by defining a {\em bilingualism threshold} $r$ beyond which a speaker would be considered bilingual. For example, setting $r=20$ every speaker who uses Catalan more than $80\%$ of time during a day is considered a Catalan monolingual, every speaker who uses Catalan less than $20\%$ is considered a Spanish monolingual, and every speaker employing both Catalan and Spanish within $20-80\%$ is considered bilingual. There is not a clear criterion about what threshold to use, so we conducted our analysis for all possible integer values of $r \in [1, 50]$. This covers all cases from the extreme in which anyone employing both languages is considered bilingual, to the stringent situation in which only those using both tongues half of the time score as non-monolinguals. 

      For each value of the bilingualism threshold we derive a full data series with a proportion $x$ of Spanish monolinguals, a proportion $b$ of bilinguals, and a proportion $y$ of Catalan monolinguals over time (Fig. \ref{fig:01}{\bf b}). We built these data series for each EULP survey and combinations of them. Most analysis were conducted on all available data; here we present results for the more robust (less noisy) time series. (See Supplementary Information for details and to explore all existing results, which are consistent throughout.) 

    \subsection{Data Availability Statement}

      The data used in this study has been keenly provided by the Institut d'Estad\'istica de Catalunya (IDESCAT) and is available in their website (http://www.idescat.cat/en/). It can also be found in the different works referenced in this paper. Notwithstanding, the data is only available at those original sources. We do not own the original data and we have not been given permission to make public a centralized summary of it. 

    \subsection{Model}

      To characterize our data we use the model by Mira et al. \cite{MiraParedes2005, MiraNieto2011, OteroMira2013, SeoaneMira2017} which considers the existence of $X$ (in this case Spanish) and $Y$ (Catalan) monolingual groups and a bilingual group $B$. These groups present fractions $x$, $y$, and $b$ of speakers respectively within a normalized population ($x + b + y = 1$). The model assumes that the probability that monolingual speakers acquire the opposite language is proportional to the prestige ($s_X$ or $s_Y$) of the other language and to the population speaking that other tongue. It is taken $s_X, s_Y \in [0, 1]$ and $s_X + s_Y = 1$ so we can focus on $s \equiv s_X$. Of all speakers acquiring a new tongue, a fraction $k$ of them retains the old one (hence becoming bilinguals) while $1-k$ of them switch and forget. The parameter $k$ is termed {\em interlinguistic similarity} \cite{MiraParedes2005, MiraNieto2011} and measures how close is the couple of languages as perceived by the population. The probabilities of leaving or entering each group ($X$, $Y$, or $B$) result in a set of differential equations that tell us the time evolution of the linguistic population:
        \begin{eqnarray}
          {dx \over dt} & = & c\Big[ (b + y)(1-k)s(1-y)^a \nonumber \\ 
          && - x\left( (1-k)(1-s)(1-x)^a + k(1-s)(1-x)^a \right) \Big], \nonumber \\
          {dy \over dt} & = & c\Big[ (b + x)(1-k)(1-s)(1-x)^a \nonumber \\ 
          && - y\left( (1-k)s(1-y)^a + ks(1-y)^a \right) \Big]. 
          \label{eq:01}
        \end{eqnarray}
      (Only two equations are needed thanks to the normalized population. Also, this is a compact version of the equations, for detailed discussion, e.g. explicit flow between groups, see \cite{MiraParedes2005, MiraNieto2011, OteroMira2013, SeoaneMira2017}.) The parameter $a$ (which has been referred to as {\em volatility} \cite{CastelloSan2013}) affects those speakers that promote language shift (termed {\em attracting population} \cite{SeoaneMira2017, HeinsaluLeonard2014}). It confers an idea of how persistent the linguistic groups are: the lower $a$ the easier it is for all groups to lose speakers, thus rendering the system more {\em volatile} \cite{ColucciOtero2014}. 

      The stability of this model has been thoroughly characterized as a function of its parameters \cite{MiraNieto2011, OteroMira2013, ColucciOtero2014, SeoaneMira2017}. If $a>1$, stable solutions include scenarios in which i) the bilinguals and either one of the monolingual groups get extinct and ii) both monolingual groups survive along a bilingual group. Coexistence is usually reached for larger $k$ and relatively balanced prestiges $s_X \sim s_Y$. 

      Equations \eqref{eq:01} are a generalization of the seminal Abrams-Strogatz model \cite{AbramsStrogatz2003} that promoted non-linear differential equations for the study of language population dynamics (even if earlier, similar approaches existed \cite{BaggsFreedman1990, BaggsFreedman1993}). The original equations did not include bilingualism on the grounds that it played a minor role for the languages under research. This is not the case in the Catalan-Spanish coexistence scenario.

      Other valuable models consider bilingual situations \cite{BaggsFreedman1990, BaggsFreedman1993, ZhangGong2013, MinettWang2008, HeinsaluLeonard2014}. Besides our familiarity with the chosen system of equations, the stability of the alternatives has not always been studied. Some of these models do not contemplate stable, coexist languages \cite{MinettWang2008} or do so only after alternative parameterizations are included \cite{BaggsFreedman1993}. It is intensely debated whether languages can coexist steadily in an asymptotic time, but it does not seem appropriate to barren that possibility beforehand. Hence, we decided to conduct our analysis with equations that allow this scenario explicitly. 

      This model consists of two coupled, non-lineal differential equations with $4$ parameters $\{a, c, k, s\}$ and two initial conditions ($x(t=t^0)$, $y(t=t^0)$). To extract these parameters from the data we followed the fitting procedure described in the Supplementary Information, which basically makes a fast, heuristic least square minimization. Also in the Supplementary Information we compare the best and worst fits and provide plots of the fits from all data series and for all bilingualism thresholds. One example of a good fit is that obtained for $r=10$, shown in Fig. \ref{fig:01}{\bf c}, along with an extrapolation towards the end of the XXI century.

      \begin{figure*}
        \centerline{\includegraphics[width=\linewidth]{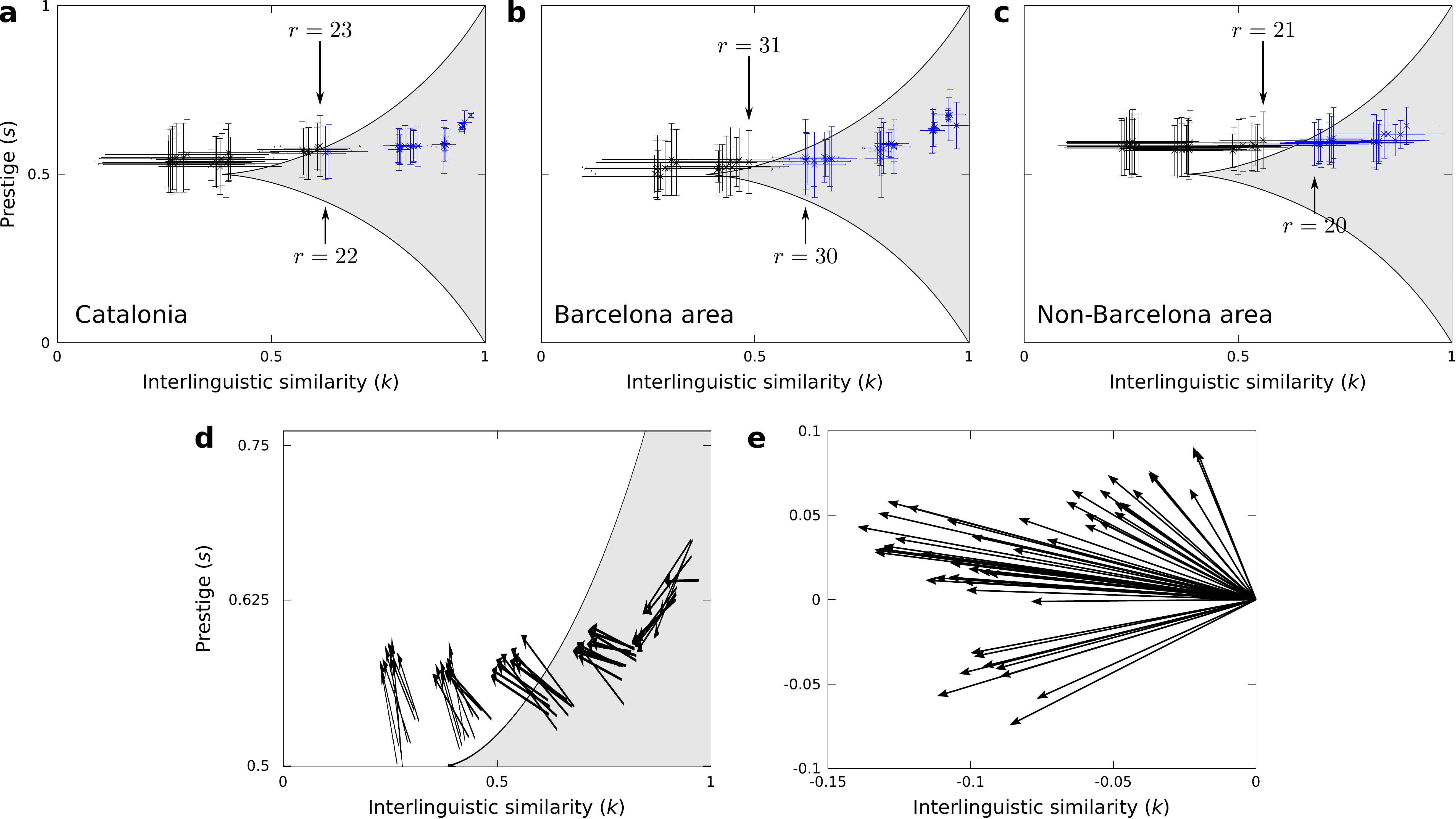}}
        \caption{{\bf Stability analysis. } Mapping measured parameters into the $k-s$ plane for {\bf a} the whole Catalan territory, {\bf b} Barcelona and metropolitan area, and {\bf c} not Barcelona. Falling outside the shaded region implies that one of the languages will go extinct (shaded region is consistent with, but does not imply, sustained coexistence). Each point represents average $(k, s)$ from the fits with a given $r$. Error bars represent one standard deviation. {\bf a} For every $r \ge 23$ the parameters imply the extinction of one of the languages. The slightly larger prestige for Spanish ($s_S \simeq 0.57$ versus $s_C \simeq 0.43$ for Catalan -- both are averages across all values of $r$) suggests a larger survival outlook for this tongue. {\bf b} Within Barcelona both prestiges are fairly equilibrated ($s_S \simeq 0.51$ versus $s_C \simeq0.49$). For $r \ge 31$ one of the languages is predicted to go extinct, but the data and the model are often compatible with the extinction of either language. {\bf c} Spanish presents a larger prestige in the rest of the Catalan territory ($s_S \simeq 0.6$ versus $s_C \simeq0.4$). {\bf d} For each $r$, arrows start in the average $(k, s)$ for Barcelona and end in the corresponding point obtained outside Barcelona. They indicate how the rural Catalan areas fall into an extinction course more easily (for lower values of $r$). {\bf e} The origin of all arrows has been shifted to $(0,0)$. We appreciate that $k$ is always notably smaller for speakers outside Barcelona. This indicates that, outside Barcelona, Catalan and Spanish are perceived as more different from each other. }\label{fig:02}
      \end{figure*}

  \section{Results}
  	\label{sec:3}

    The most important result that we extract is that the Catalan-Spanish system of coexisting languages tends, under most circumstances analyzed, to a stable state in which both languages coexist. The data also reveals a few counterintuitive insights that we examine in the next subsections. The discussion concerns mainly the parameters $k$ and $s$ extracted from adjusting the model equations to the different datasets. We can always track down the stability of the system to these two parameters and the initial conditions. The other parameters in equations \eqref{eq:01} ($a$ and $c$) are not so determinant regarding the stability. Their trends as a function of the bilingualism threshold are discussed in the Supporting Material. 

    \subsection{Stability of the Catalan-Spanish system}

      The most relevant parameters of the model are the interlinguistic similarity ($k$) and prestige ($s$) which have intuitive interpretations owing to their roles in equations \eqref{eq:01}. Thanks to previous studies of the model \cite{MiraNieto2011, OteroMira2013, ColucciOtero2014} we know how to link these parameters to the stability of the system. Figs. \ref{fig:02}{\bf a-d} show how the $k-s$ plane is divided into two regions: a gray area where coexistence is possible (depending on the intial conditions) and a white area where coexistence is never possible. For these plots, $a = 1.31$ (a value inherited from the original Abrams-Strogatz studies \cite{AbramsStrogatz2003}). For other values of $a \le 1$ a similar division of the plane happens\cite{OteroMira2013}, and values of $(k, s)$ exist for which the dynamics are equally well explained. We could have chosen any arbitrary value $a > 1$ without losing explanatory power. Comparisons between $(k, s)$ values only make sense if $a$ is fixed. Hence, to better illustrate the results, we performed our analyses both allowing $a$ to vary and keeping it fixed at $a = 1.31$. Similar conclusions are reached in both cases (see Supplementary Information), but we focus on the fixed case now. 

      \begin{figure*}
        \centerline{\includegraphics[width=\linewidth]{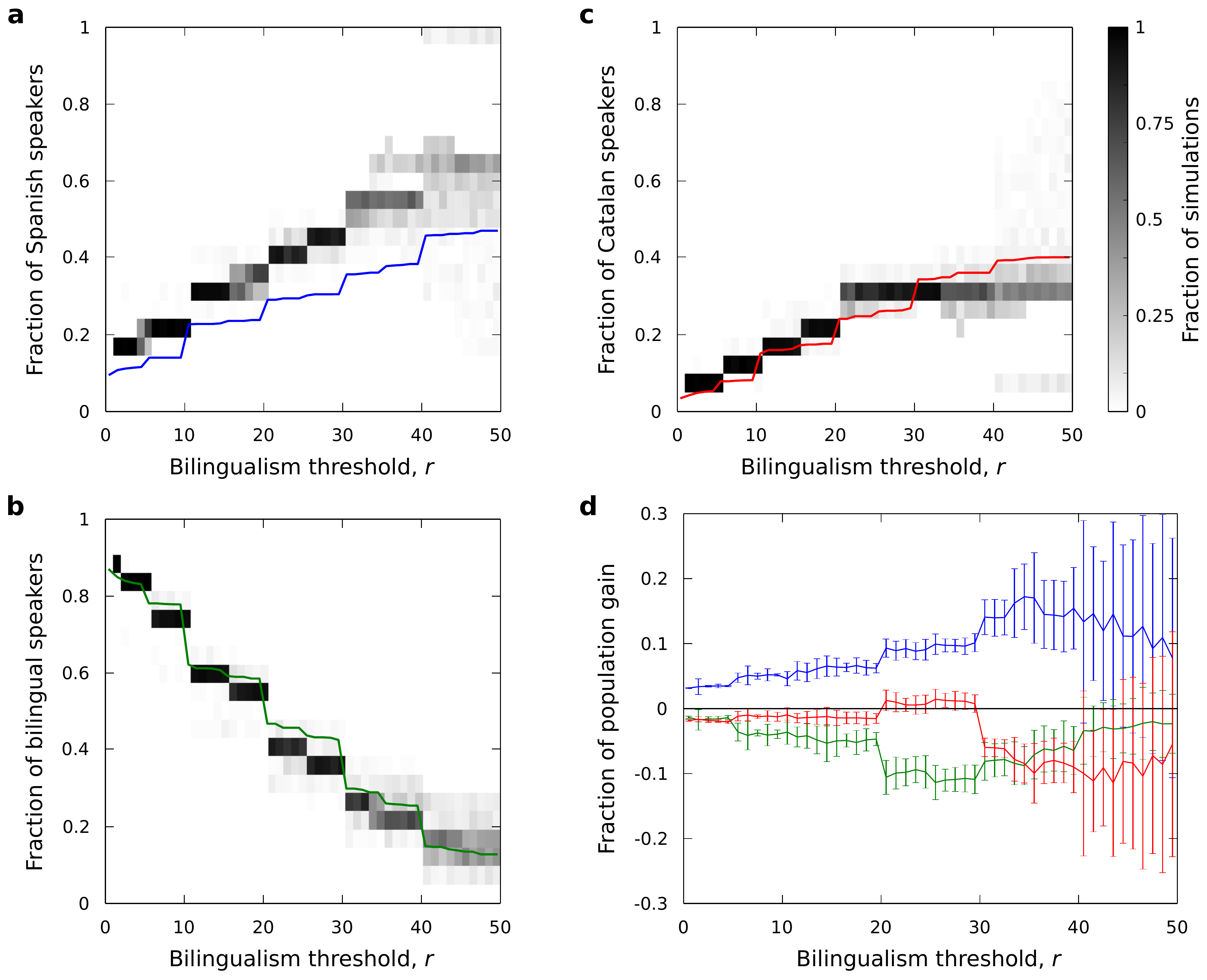}}
        \caption{{\bf Model projections for $2030$.} For each set of parameters obtained after fitting the data, equations \ref{eq:01} were evolved into the future until 2030. The fraction of speakers within each group ({\bf a}, Spanish monolinguals; {\bf b}, bilinguals; and {\bf c}, Catalan monolinguals) was then binned in $0.05$ intervals. The contributions to these intervals were weighted as indicated in the Supporting Material, so that fits with larger residua contribute less than the more accurate ones. The gray scale indicates the likelihood that a set of parameters would evolve into each $0.05$ bin for a given $r$ in $2030$. For reference, blue, green, and red lines indicate respectively the fraction of Spanish, bilingual, and Catalan speakers in $1990$, the last data point in the {\em EULP’} series. {\bf d} Average (with standard deviation) change in the fraction of speakers for each group for each $r$ value. }\label{fig:03}
      \end{figure*}

      For each data set and each bilingualism threshold we derived several collections of parameters compatible with the corresponding time series. The existence of several good fits in each case allows us to perform a statistic analysis that bootstraps the variability of the parameters. Fig. \ref{fig:02}{\bf a} shows average and standard deviations for $k$ and $s$ for the different integer values of $r \in [1, 50]$. More generous definitions of bilingualism are explained by models with larger interlinguistic similarity -- i.e. smaller values of $r$ lay at the right side of the $k-s$ plane and larger values of $r$ lay towards the middle-left. This range of $k$ values is explained by the differences in the bilingualism threshold $r$. Unluckily, the model and data available cannot offer a rigid constrain on the interlinguistic similarity. We note, though, that in average it does not become arbitrary low -- not even for the most restrictive definitions of bilingualism in which only speakers using $50\%$ of the time each language are considered bilinguals. 

      Our results also indicate that the prestige of Spanish ($s \sim 0.57$) is consistently slightly larger than that of Catalan ($1-s \sim 0.43$) for any value of $r$. These numbers are sustained throughout the whole range $r\in[1, 50]$, strongly suggesting that this is a good descriptor of the Catalan-Spanish system given the data and the model. 

      Roughly half of the $r$ values allow for asymptotic stability (up to $r \le 22$) and another half ($r \ge 23$) strictly ban it. However, for larger values of $r$ the prestige of both languages becomes more leveled ($s$ is only slightly above $0.5$ for larger $r$) and the fits become less consistent: some of them predict the extinction of one tongue and some others predict the extinction of the other one -- both usually in an asymptotic time much larger than $100$ years. This indeterminacy comes about because the system sits near a bifurcation point and the data is not enough to clarify the outcome of the competition dynamics. 

      Our stability analysis is complemented with an attempt at prediction towards $2030$. Such predictions must be taken with all the prudence possible: they are the results obtained with this model for the data available, and the seemingly open-ended nature of human dynamics does not allow us to have an all comprehensive understanding of the situation. However, some qualitative results outlined below are fairly consistent across datasets, fitting setups, and definitions of bilingualism (through the parameter $r$). This invites us to be confident about the general conclusions. 

      We registered the percentages of Spanish (Fig. \ref{fig:03}{\bf a}), Bilingual (Fig. \ref{fig:03}{\bf b}), and Catalan (Fig. \ref{fig:03}{\bf c}) speakers predicted by the model for the year $2030$ for each of the combinations of parameters derived for each time series. These year-$2030$ predictions were binned in intervals comprising $0.05$ increments in the fraction of speakers. Fig. \ref{fig:03}{a-c} shows how, consistently, our results indicate a middle-term coexistence between Spanish and Catalan, even for those configurations of parameters that imply the eventual extinction of one of the tongues. These extinction scenarios happen with stringent definitions of bilingualism (large $r$) and present relatively balanced prestiges ($s_S \sim 0.5 \sim s_C$) indicating that even if one language must die eventually, that result will only happen asymptotically and coexistence could perhaps be granted for several generations. 

      Notwithstanding the bilingualism threshold, the model always predicts a less important role for Catalan language in a middle-term future. Catalan speakers towards $2030$ would always amount to less than Spanish and, more often than not, bilingual speakers. Meanwhile, Spanish stands as the dominating language for some definitions of bilingualism, and it is consistently the group projected to grow the most until 2030. Figure \ref{fig:03}{\bf d} shows the expected gain of speakers for each group in $2030$, with Spanish standing out. For $r<30$, Spanish is expected to win most of its new speakers from bilinguals. For $r>30$ both bilinguals and Catalan monolinguals would lose a substantial amount of speakers to Spanish.

    \subsection{Analysis across different geographical areas}

      The linguistic map of Catalonia contains two very distinguishable regions: On the one hand, Barcelona and its metropolitan area constitute the second largest urban hub in Spain and agglutinates more than $70\%$ of the Catalan population. This is home to large migrant groups from the rest of Spain and elsewhere (notably Pakistan and China), while Barcelona itself is a very cosmopolitan city attracting large masses of tourists. On the other hand, the rest of Catalonia (while still containing notable urban areas and some regions with large migrant populations) has a more rural character and is spread across larger territories. 

      To further understand the linguistic reality of the system we segregated the data in those two broad geographical regions and repeated our analyses. We found similar tendencies in the parameters $a$ and $c$ (see Supplementary Information). The interlinguistic similarity again drops as the definition of bilingualism becomes more stringent (Fig. \ref{fig:02}{\bf b} and {\bf c}), again showing the limitations of the model and existing data to constrain the value of $k$. However, note once again that the interlinguistic similarity never drops to zero (not even for the most stringent definitions of bilingualism); and note also how it peaks at roughly $k=0.9$ in non-urban areas (Fig. \ref{fig:02}{\bf c}), while it comes much closer to $k=1$ in the Barcelona area, thus stressing the difference that population outside Barcelona always perceives between both languages. On the other hand, average values of prestige ($s$) are broadly consistent throughout $r\in[1, 50]$, suggesting that the model and data together are able to constrain this characteristic of the Spanish-Catalan dynamics. 

      The analysis of the parameter $s$ offers a counterintuitive result: In general, Spanish presents a lower prestige in Barcelona and its metropolitan area than in the rest of Catalonia. Note that Catalan is less spoken in Barcelona: Over the last $100$ years it never had more speakers than Spanish (Supplementary Fig. 1), while the rest of Catalonia does present a larger body of monolingual Catalan speakers (Supplementary Fig. 2). This would naively suggest that the perceived prestige of Catalan is lower in the urban metropolis (hence Spanish prestige would be higher), but our analysis indicates exactly the opposite: rural areas (where Spanish is less spoken) perceive Spanish as a more prestigious tongue. The catch is that while the decay of Spanish and Catalan in the Barcelona area has been roughly symmetric over the last century and favors a strong bilingual group (Supplementary Fig. 1{\bf a}); in regions outside Barcelona, Spanish speakers have remained relatively constant across time while the larger Catalan group has been decaying in favor of bilinguals (Supplementary Fig. 2{\bf a}). In equations \eqref{eq:01}, a large prestige captures precisely the ability of a smaller group to make a larger (and originally stronger) one decline. 

      Figs. \ref{fig:02}{\bf d-e} summarize the differences between the perception that the two geographical areas have about Catalan and Spanish. The arrows in Fig. \ref{fig:02}{\bf d} connect the averages $k$ and $s$ in Barcelona with the averages outside Barcelona. These arrows are replotted with their origin in $(0, 0)$ (Fig. \ref{fig:02}{\bf e}) to indicate how speakers outside Barcelona not only assign a slightly lower status to Catalan, but also they perceive both languages as more different -- as indicated by the $\sim 0.1$ drop in interlinguistic similarity consistent across similar definitions of bilingualism. 

      As we did for the aggregated data, we complemented the stability analysis by projecting the evolution of the model into the future until $2030$. Again, both for the metropolitan and non-metropolitan areas, most configurations of the model are compatible with middle-term coexistence of the tongues. When the definition of bilingualism is more rigorous (larger $r$) the asymmetries (either in prestige or initial conditions) become relevant and, in some cases, are capable of substantial gains and losses in number of speakers within the projected time. Within Barcelona, Catalan would be the endangered language (Supplementary Fig. 3). For almost every definition of bilingualism, Spanish would draw most of its new speakers right away from Catalan monolinguals (Supplementary Fig. 3{\bf d}). 

      Counter intuitively again, outside the metropolis, for large $r$ Catalan would be able to gain a large number of speakers from Spanish (Supplementary Fig. 4). Despite the larger prestige of Spanish in those regions, this would be possible due to the still large Catalan monolingual population outside the Barcelona area. For very low $r$ in regions outside Barcelona, the large monolingual support of Catalan would fade as the projections predict a larger presence of Spanish monolingual speakers. For intermediate $r$, Catalan would grow notably, but extracting speakers from the bilingual group rather than from the Spanish monolinguals (Supplementary Fig. 4{\bf d}).

  \section{Discussion}
  	\label{sec:4}

    In this paper we analyzed the system of Spanish and Catalan coexistence using recent and thorough data surveys and up-to-date models based on non-linear equations. There is a gap between the theoretical developments and the empirical data available \cite{SeoaneMira2017}. The former often rely on concepts (e.g. {\em bilingual speakers}) that have a clear definition within the model but that are difficult to pin down empirically. Given the complex and subjective nature of the problem under research it is necessary to rely on the self-assessment of linguistic qualities -- in this case, percentage of language use. 

    We wanted to conduct our analysis in the most general way possible given the data. We considered a series of {\em bilingualism thresholds} (encoded by $r\in[1, 50]$) and did not assume that any of these thresholds constitutes {\em the right definition of bilingualism}. Instead, we performed our analysis for all possible scenarios. This could potentially produce a wealth of models with antagonistic predictions, hence frustrating any robust conclusion. This happens often in complex systems that sit midway between competing forces -- which is the case here. Instead, our analysis renders a consistent picture across different data sources and for most different definitions of the bilingual group. This picture is that of a long term coexistence between Spanish and Catalan in Catalonia, always along a bilingual group, and with a dominating role for Spanish while the group of monolingual Catalan speakers declines lightly. 

    Further details hinge on the bilingualism threshold employed. For roughly half of the definitions of bilingualism ($r \le 22$) it is predicted that both languages will survive. The monolingual Catalan group is expected to be smaller than the Spanish one towards 2030 (Fig. \ref{fig:03}), disregarding of $r$. Most of the population would be bilingual for $r \le 22$. For very stringent definitions of bilingualism ($r>40$) some of the models predict the a huge loss of Catalan speakers before 2030, but coexistence between both languages is still the most common outcome. Similar results are obtained when segregating the data between Barcelona plus metropolitan area versus rest of Catalonia. Coexistence of both tongues remains the most persistent outcome, but for strict definitions of bilingualism ($r>40$) large losses of Catalan speakers in Barcelona and of Spanish speakers outside Barcelona become are likely within a few decades. 

    These mid-term predictions leave considerable room for action. Consequently, they are also daring and should be subjected to continuous revision as new data becomes available. The correctness of our analyses relies on some assumptions: i) The data collected so far is reliable and significant about how the situation might evolve. Consequently, ii) social and political circumstances shall not vary considerably in the future. Unluckily there are no studies about how notable political events affect the smooth dynamics of the system. (This is also true for all other models of language shift \cite{BaggsFreedman1990, BaggsFreedman1993, AbramsStrogatz2003, MiraParedes2005, Kandler2008, Castellano2009, PatriarcaHeinsalu2009, CastelloSan2013, ZhangGong2013}.) Should the socio-political stage change drastically (e.g. if Catalonia would become an independent state, a possibility debated nowadays), our analysis might become outdated. iii) We never assume that we are using {\em the right theory}. Our equations might not be correct, so indeed this exercise should help us validate the model -- even if some predictions lay far in the future. 

    In this paper we also quantified the perceived prestige of both languages and their interlinguistic similarity. The former cannot be hugely constrained by the model since we lack a definitive measure of bilingualism and, for this dataset, $k$ changes widely with $r$. Results for the prestige parameter were relatively consistent across $r\in[1, 50]$ suggesting that we have successfully captured $s$ for the languages involved. Also, spatially segregated data reveals interesting differences across regions -- notably the higher prestige of Spanish in the areas that, historically, had more Catalan speakers. In those regions also the perceived difference across languages is larger. Both these observations hold despite the variation of $r$, strongly suggesting that they are real features of the system. 

    It could be thought that an objection to our model and others similar is that the parameters are abstract and difficult to relate to more concrete features. Notwithstanding our abstractions, the parameters $s$ and $k$ have definite causal consequences in terms of population flows in our equations. Both steady states and the dynamical unfolding of the equations are intimately linked to their numerical values. Measures under different circumstances (e.g. values of $r$, or geographically stratified data) can be compared to each other and sound conclusions can be extracted. In this sense, both $k$, $s$, and other parameters carry meaningful information about the Catalan-Spanish system. We assess these quantities indirectly (by fitting the data to our model), but perceived prestige or similarity between tongues and other sociolinguistic characteristics can be directly reported in future surveys. These more concrete quantities can then be correlated to our parameters, thus helping us bridge the gap between theory and empirical data in the social sciences. 

    Mathematical models of language contact situations give us hints about the important factors that could reverse the current predictions -- notably, the perception of bilingualism and the geographic distribution of the population \cite{SeoaneMira2017}. The former is clear from our analysis and has been discussed theoretically by Heinsalu et al. \cite{HeinsaluLeonard2014}. A key for stability is thus bolstering a strong bilingual group capable of capturing speakers faster than any monolingual group. To achieve this, it is relevant that bilingual individuals reach a preponderant role within their society. Failing to establish a lasting bilingual group guarantees that the competition will result in an extinct language; and the most likely scenario, given the data, would be the decline of Catalan.

  \section*{Acknowledgments}

    We are very thankful to the Institut d'Estad\'istica de Catalunya (IDESCAT), for its collaboration supplying the data that made possible this research. We also acknowledge the attention of the Xarxa Cruscat, that responded effectively to our demands of information concerning sociolinguistic aspects of Catalonia. This work was partially funded by the Galician Royal Academy ({\em Real Academia Galega}), to which we also acknowledge a lasting insfrastructural support. Seoane wishes to acknowledge the members of the Complex Systems Lab at the Pompeu Fabra University (especially Prof. Ricard Sol\'e and Dr. Salvador Dur\'an) for useful comments and discussion. 

  \section*{Author contributions}

    All authors participated in the writing of this document. L.F.S. and J.M. coordinated the research and decided what mathematical analyses to conduct. L.F.S. implemented the mathematical and computational aspects of these analyses, elaborated the figures, and designed and implemented the fitting technique. X.L. curated the data and performed preliminary statistical analysis on it. H.M. provided qualitative knowledge about the linguistic reality of the system under research.

\end{document}